\documentstyle[aps,multicol,epsfig]{revtex}
\begin{document}
\draft
\title{Secure quantum key distribution with an uncharacterized
source}
\author{Masato Koashi$^1$ and John Preskill$^2$}
\address{$^1$CREST Research Team for Interacting Carrier Electronics,
School of
Advanced Sciences, \\
~The Graduate University for Advanced Studies (SOKENDAI),
Hayama, Kanagawa, 240-0193, Japan\\
$^2$ Institute for Quantum Information, California Institute of Technology, 
Pasadena, CA 91125, USA}

\maketitle
\begin{abstract}

We prove the security of the Bennett-Brassard (BB84) quantum key distribution protocol 
for an arbitrary source whose averaged states are basis-independent, a condition that is automatically satisfied if the source is suitably designed. The proof is based on the observation that, to an adversary, the key extraction process is equivalent to a measurement in the $\hat\sigma_x$-basis performed on a pure $\hat\sigma_z$-basis eigenstate. The dependence of the achievable key length on the bit error rate is the same as that established by Shor and Preskill for a perfect source, indicating that the defects in the source are efficiently detected by the protocol.

\end{abstract}
 
\begin{multicols}{2}

Quantum key distribution is an ingenious application of 
quantum mechanics, in which two remote parties (Alice and Bob) establish a shared secret key through the transmission of quantum signals. In the BB84 protocol \cite{BennettBrassard84}, Alice sends a key bit to Bob by preparing a qubit in one of two conjugate bases and Bob measures the qubit in one of the two bases; the eavesdropper Eve, who does not know the basis chosen by Alice or by Bob, cannot collect information about the key without producing a detectable disturbance. This protocol, when suitably augmented by classical error correction and privacy amplification, is provably secure against any attack by Eve allowed by the laws of quantum physics \cite{Mayers96,LoChau99,BBBMR00,ShorPreskill00}.

Though security can be proven without imposing any restriction on Eve's attack (other than the requirement that she has no {\it a priori} information about the basis used), it is necessary to place conditions on the performance of the source and detector employed in the protocol. In the Shor-Preskill proof \cite{ShorPreskill00}, it is assumed that any flaws in the source and detector can be absorbed into Eve's basis-independent attack. The proof by Mayers \cite{Mayers96}, however, applies to a more general setting: although the source is perfect, the detector has never been tested and is completely {\em uncharacterized}. Indeed, the detector could be under the control of Eve's collaborator Fred. Fred is unable to send messages to Eve, but he knows Bob's basis and can adjust the measurement performed by the detector accordingly. Still, as Mayers showed, Fred cannot fool Alice and Bob into accepting a key that Eve knows, as long as the efficiency of the detector is basis independent. Since a real device could have an indefinite number of degrees of freedom, no test can fully characterize it; therefore proving security in the case of an uncharacterized apparatus provides comfort to a highly suspicious user of the key distribution scheme.

In this Letter, we present a simple proof of the security of the BB84 protocol that applies to a setting opposite to that considered by Mayers: the detector is perfect and Fred controls the {\em source}. We will, however, place one important restriction on Fred's attack --- the source must not leak any information to Eve about the basis chosen by Alice. That is, the state emitted by the source, averaged over the values of Alice's key bit, is required to be independent of Alice's basis. Our proof applies to faulty sources that are notably more general than those encompassed by the Shor-Preskill proof; to give just one example, it applies to a source that performs perfectly when Alice chooses the $\hat\sigma_x$-basis but that rotates the qubit when Alice chooses the $\hat\sigma_z$-basis. Nevertheless, our proof shows that secure key can be extracted from sifted key at the same rate established by Shor and Preskill.

Our proof combines insights gleaned from both the Mayers proof and the Shor-Preskill proof. Following Mayers, we analyze the information about (Bob's) key collected by Eve in the case where Alice and Bob are using {\em different} bases. Following Shor and Preskill, we bound Eve's information by observing that Bob {\em could have} performed error correction to remove any entanglement with Eve's probe before executing the measurement that extracts his final key. The core of our proof is the observation that a single quantum circuit computes Bob's final key in the $\hat\sigma_x$-basis and reverses the damage inflicted by Eve if the error rate is small in the $\hat\sigma_z$-basis. Using the same method, we can also prove security for the case of an uncharacterized detector, allowing a more general source and establishing a higher rate of key generation than in the proof by Mayers.

Before proceeding to the proof, let us specify in more detail our models of the source and detector. Alice prepares a physical system with Hilbert space ${\cal H}_{\rm A}$,
which has an arbitrary size, in one of four states $\hat\rho(a,g)$ with probability $p_{a,g}$;
$a=0,1$ labels Alice's basis choice and $g=0,1$ is the value of her key bit. The choice of $a$ is assumed to be completely random: $p_{a,0}+p_{a,1}=1/2$. We assume that the states satisfy 
\begin{equation}
p_{0,0}~\hat\rho(0,0)+p_{0,1}~\hat\rho(0,1)
=p_{1,0}~\hat\rho(1,0)+p_{1,1}~\hat\rho(1,1),
\label{source-cond}
\end{equation}
which is vital in the security proof. A convenient way to prepare such an ensemble is to introduce an auxiliary system A$^\prime$ with Hilbert space ${\cal H}_{\rm A'}$. Alice first prepares ${\cal H}_{\rm A}\otimes {\cal H}_{\rm A'}$ in an entangled state $\hat\rho_{\rm AA'}$, and then performs a measurement $M_{a}$ on system A$^\prime$ alone. The measurement $M_{a}$ gives a binary outcome, determining $g$. Eq.~(\ref{source-cond}) is then satisfied because the choice of the measurement, $M_{0}$ or $M_{1}$, does not affect the marginal state of 
${\cal H}_{\rm A}$. Hence, if the source is realized in this way, there is no need to carry out tests to characterize its performance. 

As noted in \cite{GottesmanPreskill01}, if ${\rm A}'$ is a qubit, $M_0$ is a measurement of $\hat\sigma_z$, and $M_1$ is a measurement of $\hat\sigma_x$, then security can be established by the method of Shor and Preskill. But our security proof invokes only the condition (\ref{source-cond}); no further properties of $\hat\rho_{\rm AA'}$, $M_{0}$, or $M_{1}$ need be specified. 

At the end of the transmission channel 
${\cal H}_{\rm A} \rightarrow {\cal H}_{\rm B}$, Bob switches between two
measurements on ${\cal H}_{\rm B}$.
 We assume that the two measurements
are modeled by a common quantum channel ${\cal H}_{\rm B}\rightarrow {\cal
H}_2$, where $\dim{\cal H}_2=2$, followed by the measurement of 
the Pauli operator $\hat\sigma_z$ or $\hat\sigma_x$. 
In the security proof, we include the common 
quantum channel in the transmission channel between 
Alice and Bob, so that Bob receives a qubit at the end of 
the channel.

The protocol that we shall prove to be secure is the following:
Let $\Omega\equiv\{1,\ldots,4N(1+\epsilon)\}$.
The variable denoted by $\bar{a}$ takes the value opposite to 
$a$. 

{\em Protocol 1 (BB84)} --- 
(1) Alice creates
 random bit sequences $\{a_i\}$ and $\{g_i\}$ for $i\in \Omega$.
Alice randomly chooses  
a subset $R\subset\Omega$ with size 
$|R|=2N(1+\epsilon)$.
(2) Bob creates
a random bit sequence $\{b_i\}$.
(3) When $i\in R$, Alice sends $\hat\rho(a_i,g_i)$.
When $i\in \bar{R}(\equiv \Omega-R)$, Alice sends $\hat\rho(\bar{a}_i,g_i)$.
(4) Bob measures $\hat\sigma_z$ when $b_i=0$,
and measures $\hat\sigma_x$ when $b_i=1$. For either case, 
 he sets bit $h_i$ according to the outcome ($h_i=0$ for outcome $1$ and
$h_i=1$ for outcome $-1$).
(5) Bob announces $\{b_i\}$. Alice announces $\{a_i\}$ and $R$.
If the size of $T\equiv \{i\in R|a_i=b_i\}$ is less than 
$N$, the protocol aborts. Bob decides randomly on
a subset $S\subset \{i\in \bar{R} | \bar{a}_i = b_i\}$
with $|S|=N$ and announces (if he cannot do this, the protocol aborts). 
(6) Alice and Bob compare $g_i$ and $h_i$ for $i\in T$ and 
determine the error rate $\delta$. If $\delta$ is too large,
the protocol aborts.
(7) Bob randomizes the positions 
of the $N$ qubits in $S$ by a permutation $\pi$ and announces 
$\pi$. Bob announces a linear code
$C$ with $|C|=2^r$ that corrects $N(\delta+\epsilon)$ errors 
occurring in random positions with probability 
exponentially close to unity.
(8) The sifted key $\kappa_{\rm sif}$
of length $N$ is defined as the sequence $\{h_i\}_{i\in S}$.
The final key is the coset $\kappa_{\rm sif}+C^\perp$.
(9) Alice obtains $\kappa_{\rm sif}$ by applying an
error correction scheme to $\{g_i\}_{i\in S}$ via
encrypted communication with Bob, consuming $\tau$ bits
of the previously shared secret key. Then Alice obtains the final key.

Protocol 1 is the standard BB84 protocol, except for the use of $\bar a_i$ in place of $a_i$ in steps (3) and (5), which we have adopted for later convenience in the proof.
The random permutation $\pi$ in step (7) is redundant, since it suffices to choose the code $C$ randomly instead of doing the permutation.
In the limit of large $N$,  the achievable $r/N$ reaches 
$1-h(\delta)$ (where $h(\delta)=-\delta\log_2\delta - (1-\delta)\log_2(1-\delta)$ is the binary entropy function), and $\tau/N$ in step (9) approaches
$h(\delta)$, resulting in the rate of key generation 
$1-2h(\delta)$.

Our proof uses some basic properties of (classical) error-correcting codes. 
The linear code $C$
appearing in step (7) is an
$r$-dimensional subspace of the binary vector space ${\bf F}_2^N$.
The code $C^\perp$ appearing in step (8) is 
the orthogonal complement of $C$, called the dual of $C$.
We can specify a linear coding function 
$G:{\bf F}_2^r \rightarrow {\bf F}_2^N$, which assigns a distinct codeword
of $C$ to each binary sequence of length $r$.
We have assumed in the protocol that 
$C$ corrects $N(\delta+\epsilon)$ errors 
occurring in random positions with probability 
exponentially close to unity. More specifically,
there exists a set of correctable errors ${\cal E}\subset{\bf F}_2^N$ and 
a decoding function $f:{\bf F}_2^N \rightarrow {\bf F}_2^r$,
satisfying 
\begin{equation}
f(G(y)+x)=y
\label{decoding}
\end{equation}
for any $y\in {\bf F}_2^r$ and any $x \in {\cal E}$.
A random error with weight at most $N(\delta+\epsilon)$
belongs to 
${\cal E}$ with probability 
exponentially close to unity.
The function $f$ is not necessarily linear and may be hard to
compute, but we will need only its existence for the proof of 
security --- Bob does not compute $f$ in the actual protocol.

What Bob actually calculates is the 
coset $\kappa_{\rm sif}+C^\perp$ in step (8). One way to do this 
is to use the function $G^T:{\bf F}_2^N \rightarrow {\bf F}_2^r$,
which is the adjoint (matrix transpose) of $G$ 
satisfying $G^T(x) \cdot y=x \cdot G(y)~ ({\rm mod} ~2)$ for any
$x\in {\bf F}_2^N$ and $y \in {\bf F}_2^r$.
Since the kernel of $G^T(x)$ is $C^\perp$, the final key 
is the $r$-bit sequence $G^T(\kappa_{\rm sif})$.
The duality between $G$ and $G^T$ will play an important role
in the security proof below.

In order to prove that protocol 1 is secure, we need to show that Eve's 
maximum knowledge $I_1$
about the final key is negligible. Note that Bob's final key 
is determined at step (8); step (9), which assures that Alice's key agrees with Bob's and leaks no information to Eve, is not relevant to $I_1$. 
Let us compare Protocol 1 with a modified one: 

{\em Protocol 2} --- (3)$^\prime$ When $i\in R$, Alice sends
$\hat\rho(a_i,g_i)$. When $i\in \bar{R}$, Alice also sends $\hat\rho(a_i,g_i)$.
The other steps are the same as Protocol 1. 

This modification follows Mayers's argument
\cite{Mayers96} except for the exchanged roles of the sender 
and the receiver. The only difference between protocols 1 and 2 is a flip in Alice's basis for $i\in \bar{R}$. But the bits $\{g_i\}$ for $i\in \bar{R}$ are kept secret by Alice. Hence, for Eve and Bob only the state averaged over $\{g_i\}$ is relevant, and this state is identical for the two protocols by the condition Eq. (\ref{source-cond}). Therefore, Eve's maximum knowledge $I_2$ about Bob's final key in Protocol 2 is the same as $I_1$.

Next, let us further modify Protocol 2 in favor of Eve, by allowing Eve to control Alice's source. Now Eve knows $\{a_i\}$ and $\{g_i\}$ and is free to prepare the states measured by Bob however she pleases. Since the states $\hat\rho(a_i,g_i)$ have been removed from the protocol and Bob's measurements are symmetric in $b_i$, the protocol is completely symmetric in $\{a_i\}$ and $\{g_i\}$. Therefore we may assume $a_i=g_i=0$ without 
loss of generality. The resulting protocol is as follows:

{\em Protocol 3} --- 
(1) Alice randomly chooses  
a subset $R\subset\Omega$ with size 
$|R|=2N(1+\epsilon)$.
(2) Bob creates
a random bit sequence $\{b_i\}$.
(3) Eve prepares Bob's qubits and her ancilla system in a state.
(4) Bob measures $\hat\sigma_z$ when $b_i=0$,
and measures $\hat\sigma_x$ when $b_i=1$. For either case, he sets the bit $h_i$ according to the outcome.
(5) Bob announces $\{b_i\}$. Alice announces $R$.
If the size of $T\equiv \{i\in R|b_i=0\}$ is less than 
$N$, the protocol aborts. Bob decides randomly on
a subset $S\subset \{i\in \bar{R} |b_i=1\}$
with $|S|=N$ and announces (if he cannot do this, the protocol aborts). 
(6) Bob counts the number $n$ of bits with $h_i=1$ for $i\in T$ and 
determines the error rate $\delta=n/|T|$. If $\delta$ is too large,
the protocol aborts.
(7) Bob randomizes the positions 
of the $N$ qubits in $S$ by a permutation $\pi$ and announces 
$\pi$. Bob announces a linear code $C$.
(8) The sifted key $\kappa_{\rm sif}$
of length $N$ is defined as $\{h_i\}_{i\in S}$.
The final key is the coset $\kappa_{\rm sif}+C^\perp$.

Since the modifications in the protocol favor Eve, Eve's maximum knowledge $I_3$ about the final key in Protocol 3 is no less than $I_2$: thus $I_1=I_2\le I_3$. To complete the proof, we will show that $I_3$ is small --- Eve cannot predict Bob's key accurately because Bob is measuring in the ``wrong'' basis. 

Let us denote the Hilbert space of Eve's system as ${\cal H}_{\rm E}$ and that of the $N$ qubits belonging to $S$ as ${\cal H}_S$. We may imagine that Bob's measurement on set $S$ is delayed until  step (8), and denote by $\hat\rho$ the state over ${\cal H}_S\otimes {\cal H}_{\rm E}$ after the verification test on the set $T$ is done, but before the qubits in $S$ are measured. The test on $T$ finds that the rate of error ($\hat{\sigma}_z=-1$) over $N$ (or more) randomly chosen qubits is $\delta$. If the qubits in the set $S$ were also measured in the $\sigma_z$-basis, then the joint probability of finding an error rate less than $\delta$
in $T$ and finding more than $N(\delta+\epsilon)$ errors in $S$ would be
asymptotically less than $\exp[-\epsilon^2 N/ 4(\delta-\delta^2)]$
for any strategy by Eve. Ignoring any inefficient strategy that has only an exponentially small probability of giving an error rate less than $\delta$ in $T$, we conclude that for the state $\hat\rho$, the probability of finding more than $N(\delta+\epsilon)$ errors in $S$ is exponentially small.

Let $\{|v\rangle_Z, v\in {\bf F}_2^N\}$ denote the ``$Z$-basis'' of ${\cal H}_S$, where the value of the $j$-th bit of $v$ corresponds to the eigenvalue of $\hat{\sigma}_z$ on 
the $j$-th qubit, and let $\{|v\rangle_X= \hat H^N|v\rangle_Z\}$ denote the ``$X$-basis,'' where $\hat H^N$ is the Hadamard transformation acting on the $N$ qubits. The announcement of $\pi$ in step (7) can be described as
the transmission from Bob to Eve of a particle J in one of $N!$ orthogonal 
states $\{|\pi\rangle_{\rm J}\}$. The symmetrized state held by Bob and Eve after transmission of the particle is  
\begin{equation}
\hat{\rho}_{\rm s}=(N!)^{-1}\sum_\pi |\pi\rangle_{\rm J}\langle\pi|
\otimes (\hat{U}_\pi\otimes \hat{1}_{\rm E})\hat\rho(\hat{U}^\dagger_\pi\otimes
\hat{1}_{\rm E}).
\end{equation}
Let $\hat{P}_{\cal E}$ be the projection of ${\cal H}_S$ onto the subspace spanned by the states $|e\rangle_Z$ such that $e\in {\cal E}$. The successful verification test ensures that
the probability of finding an error pattern that is not in ${\cal E}$ is exponentially small: 
${\rm Tr}[(\hat{P}_{\cal E}\otimes\hat{1}_{\rm E})\hat{\rho}_{\rm s}]\ge 1-\eta,$
where $\eta$ is an exponentially small number. (We are now regarding the particle J as part of Eve's system E.) If we define $\hat{\rho}'$ as 
\begin{equation}
\hat{\rho}'\equiv \frac{(\hat{P}_{\cal E}\otimes\hat{1}_{\rm E})\hat{\rho}_{\rm s}
(\hat{P}_{\cal E}\otimes\hat{1}_{\rm E})}{{\rm
Tr}[(\hat{P}_{\cal E}\otimes\hat{1}_{\rm E})\hat{\rho}_{\rm s}]},
\label{correctablerho}
\end{equation}
its fidelity \cite{fid} to $\hat\rho_{\rm s}$,
$F(\hat{\rho}',\hat{\rho}_{\rm s})\equiv 
[{\rm Tr}(\sqrt{\hat{\rho}'}\hat{\rho}_{\rm s}\sqrt{\hat{\rho}'})^{1/2}]^2$, is
given by
\begin{equation}
F(\hat{\rho}',\hat{\rho}_{\rm s})=
{\rm Tr}[(\hat{P}_{\cal E}\otimes\hat{1}_{\rm E})\hat{\rho}_{\rm s}]\ge 1-\eta.
\end{equation}
In what follows, we will show that if the state $\hat{\rho}'$ instead of $\hat{\rho}_{\rm s}$ were used, Eve would have no information about the final key ($I_3=0$).
Then we will infer that any actual strategy by Eve (that passes the verification test with a probability that is not exponentially small) gives her exponentially small information. 

In Protocol 3, Bob measures in the $Z$-basis for the verification test, and in the $X$-basis to generate the key --- we need to show that if the error rate is low in the $Z$-basis, then the key is random and private. Our proof invokes a quantum circuit that outputs the same $r$-bit final key as Bob finds in Protocol 3, and that also expunges Eve's entanglement with the key bits. Though Bob might not have actually executed this circuit, it would be all the same to Eve if he had, which is sufficient to ensure privacy.

The circuit, shown in Fig.~1, uses an auxiliary system Q of $r$ qubits initially prepared
in the state $|0\rangle_X$, and is a composition $\hat{U}\equiv \hat{U}_2\hat{U}_1$
of two unitary operators $\hat{U}_1$ and $\hat{U}_2$. The operator $\hat{U}_1$, which calculates the final key,
acts in the $X$-basis as 
\begin{equation}
\hat{U}_1:|x\rangle_X\otimes |y\rangle_X \rightarrow
|x\rangle_X \otimes |y+G^T(x)\rangle_X.
\label{U1X}
\end{equation}
Using the duality between $G$ and $G^T$, we easily see that $U_1$ acts in the
$Z$-basis as 
\begin{equation}
\hat{U}_1:|x\rangle_Z\otimes |y\rangle_Z \rightarrow
|x+G(y)\rangle_Z \otimes |y\rangle_Z.
\label{U1Z}
\end{equation}
The operator $\hat{U}_2$ is defined in the $Z$-basis as 
\begin{equation}
\hat{U}_2:|x\rangle_Z\otimes |y\rangle_Z \rightarrow
|x\rangle_Z \otimes |y+f(x)\rangle_Z,
\label{U2Z}
\end{equation}
and in the $X$-basis acts as
\begin{equation}
\hat{U}_2:|x\rangle_X\otimes |y\rangle_X \rightarrow
|\Psi_{x,y}\rangle_X \otimes |y\rangle_X.
\label{U2X}
\end{equation}
Here $|\Psi_{x,y}\rangle$ is a rather complicated 
state of ${\cal H}_S$, but its exact form is not relevant here.

\begin{figure}
\centerline {\epsfig{width=8.0cm,file=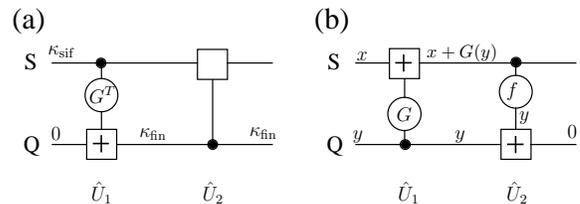}}
\caption{(a) A quantum circuit calculating 
$\kappa_{\rm fin}=G^T(\kappa_{\rm sif})$
in the $X$-basis. (b) The same circuit in the $Z$-basis. 
When $x\in {\cal E}$, the final state of system Q 
is $|0\rangle_Z$.
}
\end{figure}

If the initial state of the ancilla Q is $|0\rangle_X$, then
from Eqs.~(\ref{U1X}) and (\ref{U2X}) we have
\begin{equation}
\hat{U}(|\kappa_{\rm sif}\rangle_X\otimes |0\rangle_X)=
 |\Psi_{x,\kappa_{\rm fin}}\rangle_X \otimes|G^T(\kappa_{\rm sif}) \rangle_X;
\end{equation}
the final key $\kappa_{\rm fin}=G^T(\kappa_{\rm sif})$ is obtained 
by measuring the system Q in the $X$-basis after execution of the circuit.
On the other hand, Eqs.~(\ref{U1Z}) and (\ref{U2Z}) with $|0\rangle_X\propto\sum_y|y\rangle_Z$
lead to
\begin{eqnarray}
&&\hat{U}(|x\rangle_Z\otimes |0\rangle_X)\nonumber\\
\propto
&&\sum_y |x+G(y)\rangle_Z \otimes|y+f(x+G(y)) \rangle_Z;
\end{eqnarray}
if the initial state $|x\rangle_Z$ satisfies
$x\in {\cal E}$, the final state of Q is $|0\rangle_Z$,
due to Eq.~(\ref{decoding}). Then, Eq.~(\ref{correctablerho})
ensures that if the initial state of ${\cal H}_S\otimes {\cal H}_E$
is $\hat\rho'$, the final marginal state of Q is still $|0\rangle_Z$.
Therefore, the final state $\hat\rho_{\rm Q}$ of Q obtained 
when we start from the actual state $\hat\rho_{\rm s}$
is exponentially close to $|0\rangle_Z$:
\begin{equation}
_Z\langle 0|\hat\rho_{\rm Q}|0\rangle_Z\ge F(\hat\rho',\hat\rho_{\rm s})\ge 1-\eta.
\label{rhoRfidelity}
\end{equation}

Eq.~(\ref{rhoRfidelity}) establishes that the final key 
can be obtained from a complete $X$-basis measurement on the state $\hat{\rho}_{\rm Q}$, 
whose fidelity to the $Z$-basis eigenstate $|0\rangle_Z$ is exponentially 
close to unity. From this, we conclude the following:
(a) The mutual information $I_3$ between the final key and Eve, who may conduct any 
measurement on her system, is upper-bounded by the von Neumann entropy $S(\hat{\rho}_{\rm Q})$
\cite{Holevo73}.
Since $\hat{\rho}_{\rm Q}$ has an eigenvalue greater than or equal to 
$1-\eta$, we have $I_1\le I_3\le S(\hat{\rho}_{\rm Q})\le h(\eta)
+\eta\log_2 (2^r-1)<h(\eta)+r\eta$.
(b) The probability distribution $p_y$ over the $2^r$ final keys 
is very close to uniform. In fact, 
the fidelity to the uniform distribution cannot 
be lower than  
the fidelity in Eq.~(\ref{rhoRfidelity}). Thus we have
$2^{-r}(\sum_y \sqrt{p_y})^2\ge 1-\eta$. Using the
inequality 
$-2\log_2 x\le r(1-x)^2+(1-x^2)/(\log_e 2)$ which holds for 
$x\ge 2^{-r/2}$ when $r$ is large, 
the Shannon entropy of $\{p_y\}$ is bounded
as $H(\{p_y\})=r+2 \sum_y p_y \log_2 (2^r p_y)^{-1/2}\ge 
r(2\sqrt{1-\eta}-1) \ge r(1-2\eta)$.

The two imperfections of the final key derived in (a) and (b)
can be combined into a single parameter by the following argument.
Let us assume that Bob randomly
chooses and announces a bit sequence $w\in {\bf F}_2^r$, and 
produces a new key $w+y$ which is truly uniformly distributed. If Eve's information about $y$ is $I_1$, then her information about $w+y$ is 
\begin{equation}
I= r - H(\{p_y\}) + I_1< 3r\eta+h(\eta),
\end{equation}
which is also exponentially small, concluding the proof.

Finally, suppose that Bob uses a detector with imperfect efficiency,
which has a ``null'' outcome (signifying a detection failure) 
in addition to the valid binary outcome.
Our proof remains valid, provided that the efficiency (probability of obtaining a valid
outcome) is the same for the two bases, and the size of $\Omega$ is 
increased appropriately. 

Our proof of security applies to an uncharacterized source with basis-independent averaged states. By interchanging the roles of sender and receiver, the same proof can be applied to the case of an uncharacterized detector, considered by Mayers \cite{Mayers96}. Indeed, in that case our proof allows a more general source (one triggered by a perfect measurement on half of an entangled state, as opposed to a perfect source) and a higher rate of key generation ($1-2h(\delta)$ rather than $1-h(\delta)-h(2\delta)$) than established by Mayers. In either case, by exploiting the duality between the operation that encodes a message using $C$ and the operation that computes a $C^\perp$ coset, our proof illuminates the connection between a low error rate and successful privacy amplification.

It is also interesting to consider {\em characterized} imperfect sources and detectors that have limited basis-dependent flaws.  One important case of a characterized defective source, recently analyzed in \cite{ILM01}, is a source that occasionally emits two identical copies of a qubit, one of which can be intercepted by Eve. In this case, our proof does not apply because Eq.~(\ref{source-cond}) is not satisfied.  Security criteria for characterized sources and detectors are further discussed in \cite{GLP}.

We thank David DiVincenzo, Peter Shor, Andy Yao, and especially Daniel Gottesman and Hoi-Kwong Lo for helpful discussions.
This work has been supported in part by the Department of Energy under Grant No. DE-FG03-92-ER40701, by the National Science Foundation under Grant No. EIA-0086038, and by the Caltech MURI Center for Quantum Networks under ARO Grant No. DAAD19-00-1-0374.

\end{multicols}
\end{document}